\documentclass{PoS}
\usepackage{graphicx}

\title{Precision tests of the $J/\psi$ from full lattice QCD: mass, leptonic width and radiative decay rate to $\eta_c$}

\ShortTitle{$J/\psi$ from full lattice QCD}

\author{\speaker{C. T. H. Davies}, G. C. Donald, R. J. Dowdall, J. Koponen\\
          \\
        University of Glasgow\\
        E-mail: \email{Christine.Davies@glasgow.ac.uk}}

\author{E. Follana\\
        Universidad de Zaragoza}
\author{K. Hornbostel\\
        Southern Methodist University}
\author{G. P. Lepage\\
        Cornell University}
\author{C. McNeile\\
        Universit\"{a}t Wuppertal}
\author{HPQCD collaboration}

\abstract{We show results from calculations in full lattice QCD of 
the mass, leptonic width 
and radiative decay rate to $\eta_c$ of the $J/\psi$ meson. These provide 
few \% tests of QCD. Another (1.5\%) test comes from comparison of time-moments of the 
vector charmonium correlator with results derived from the experimental values 
of $R(e^+e^- \rightarrow \mathrm{hadrons})$ in the charm region.} 

\FullConference{Xth Quark Confinement and the Hadron Spectrum,\\
		October 8-12, 2012\\
		TUM Campus Garching, Munich, Germany}

\begin{document}

\section{Introduction}
Precision tests of lattice QCD against experiment are critical 
to provide benchmarks against which to calibrate the reliability 
of predictions from lattice QCD~\cite{original}. Many of these tests are also important 
tests of QCD itself because frequently lattice QCD provides the only 
method for precision calculation. Most tests carried out so far have 
been based on the spectrum of gold-plated hadron masses. It is important 
to have tests that also involve decay matrix elements but, for weak decays,
uncertainties in CKM elements are a limiting factor (indeed, the 
lattice calculations are used to determine the CKM elements). 
Electromagnetic decay rates therefore have the potential to provide good tests
because their normalisation, related to $\alpha_{QED}$, is well-known. 

Here we provide such tests for the $J/\psi$ including the full 
effect of $u$, $d$ and $s$ quarks in the sea for the first time. 
We are able to do this 
because we have developed a particularly accurate formalism 
for discretising 
the quark piece of the QCD Lagrangian onto a lattice, known as 
Highly Improved Staggered Quarks (HISQ)~\cite{hisq}. The work described here 
has now been published~\cite{psipaper} and we refer to that paper for all details. 
Here we simply give highlights from the results. For a similar 
calculation using twisted mass quarks and including $u$ and $d$ 
quarks in the sea, see~\cite{etm}. For earlier work in the quenched approximation see~\cite{dudek}. 

\begin{figure}
\includegraphics[width=0.5\hsize]{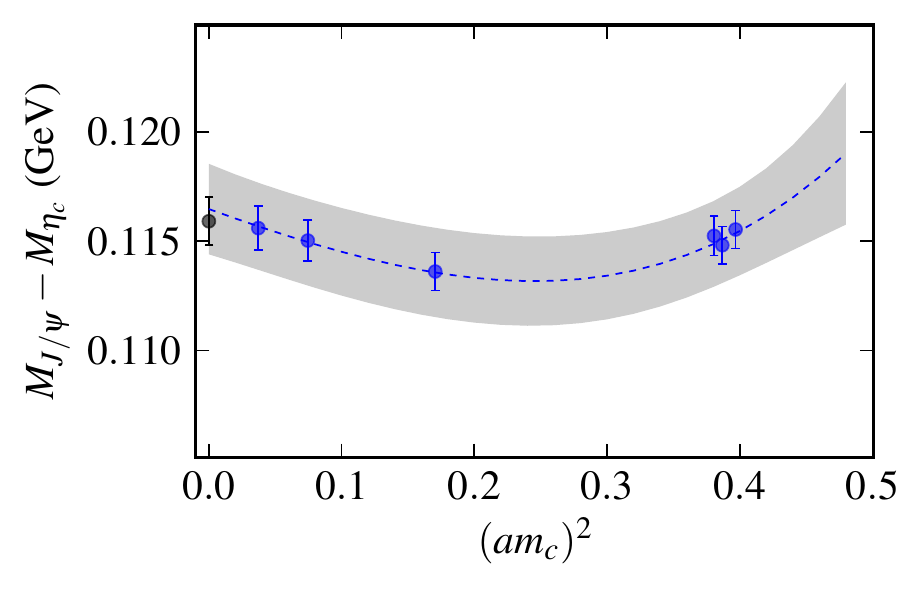}
\includegraphics[width=0.5\hsize]{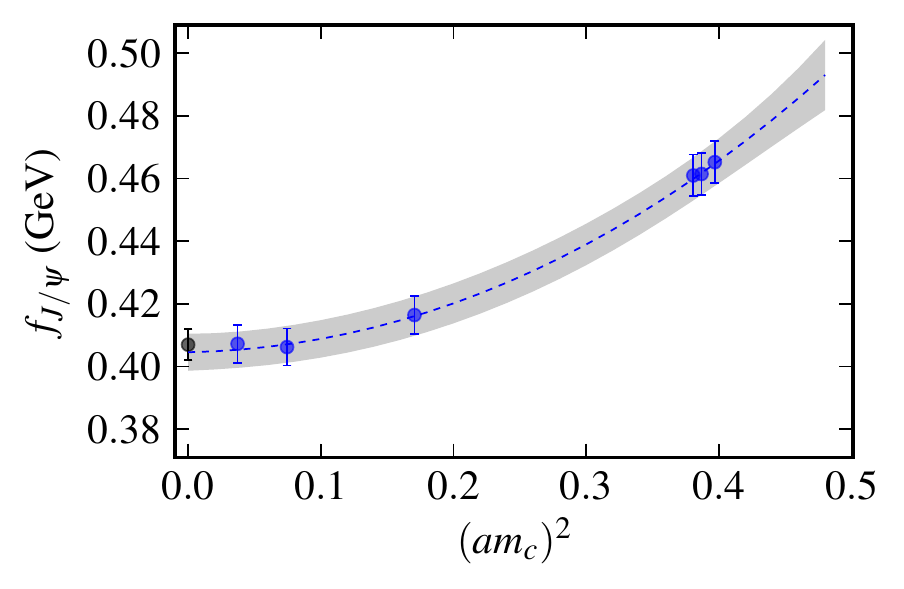}
\caption{The charmonium hyperfine splitting on the left and decay 
constant of the $J/\psi$ on the right as a function of lattice 
spacing. The grey band is our fit to the calculated blue points. 
At $a=0$ it agrees well with the experimental 
value given by the black point. }
\label{fig:hf}
\end{figure}

\section{Results}

Figure~\ref{fig:hf} shows results on the left for the hyperfine 
splitting ($M_{J/\psi}-M_{\eta_c}$) and on the right for the 
$J/\psi$ decay constant, $f_{J/\psi}$, as a function of lattice spacing, $a$ 
(given in units of the $c$ quark mass, $m_c$). The blue 
points show our results at four different widely spaced values 
of $a$ and the grey band shows our fit, including the $1\sigma$ error 
bar, which allows extrapolation 
to the real world at $a=0$. The black point at $a=0.0$ is the 
experimental value. For $f_{J/\psi}$ this is derived from 
$\Gamma(J/\psi \rightarrow e^+e^-)$. 

Our value for the hyperfine 
splitting is 116.5(2.1)(2.4) MeV where the first error is from our 
calculation and fit and the second is from uncertainties in the $\eta_c$ 
mass from electromagnetic effects and $\eta_c$ annihilation, neither 
of which are included in our lattice QCD calculation. The current 
experimental average is 115.9(1.1) MeV~\cite{pdg}. Our value for $f_{J/\psi}$ is
405(6)(2) MeV to be compared to 407(5) MeV from experiment~\cite{pdg}.   

\begin{figure}
\includegraphics[width=0.5\hsize]{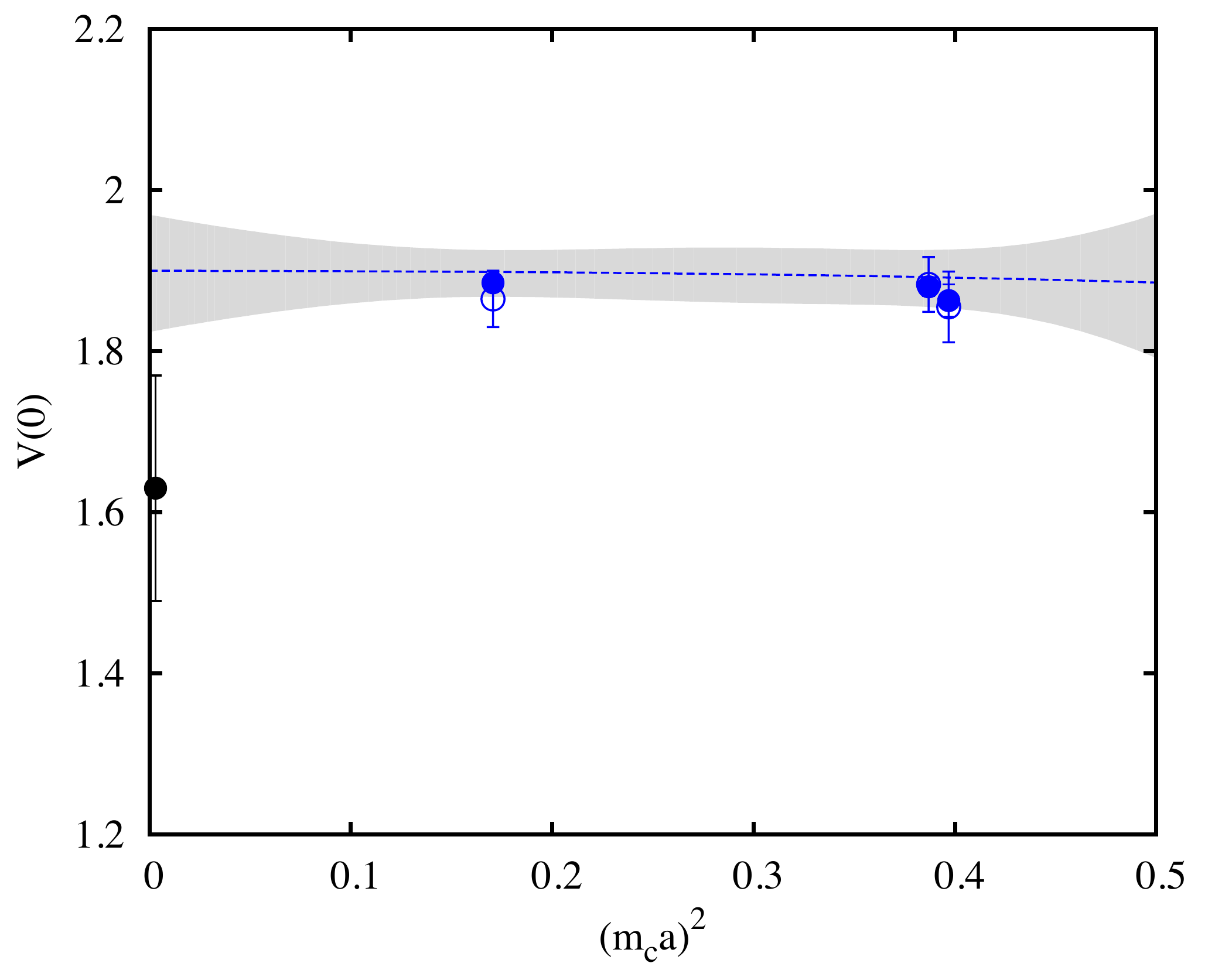}
\includegraphics[width=0.4\hsize]{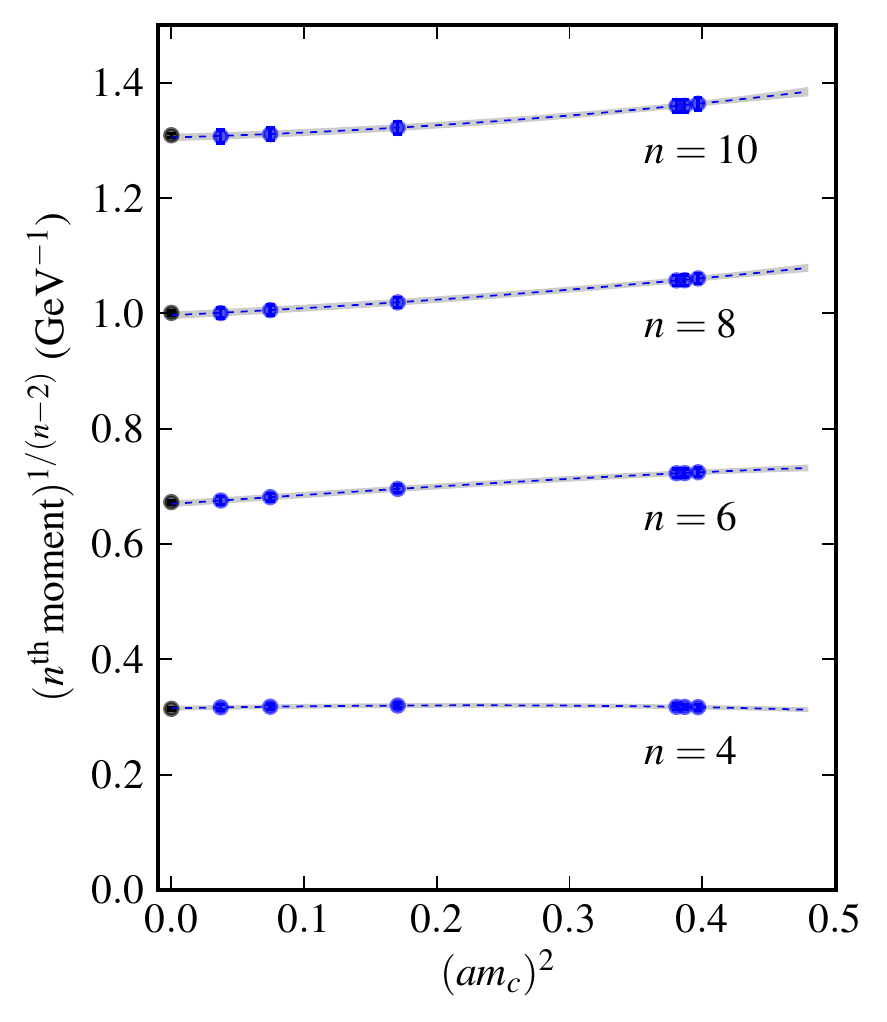}
\caption[kjh]{On the left is the value of the vector form factor 
between $J/\psi$ and $\eta_c$ at squared 4-momentum transfer equal to zero, 
and on the right the $n$th time-moments of the vector charmonium correlator
for $n=4$, 6, 8, and 10. 
Symbols are as in Figure~\ref{fig:hf}.}
\label{fig:vR}
\end{figure}

Figure~\ref{fig:vR} shows on the left results for $V(0)$, the vector form factor 
at $q^2=0$ between the $J/\psi$ and $\eta_c$, which is related the 
decay rate for $J/\psi \rightarrow \gamma \eta_c$. Our final result for 
V(0) is 1.90(7)(1) giving $\Gamma(J/\psi \rightarrow e^+e^- \gamma \eta_c)$ = 2.49(18)(7) keV. 
This is to be compared to the experimental value of 1.84(30) keV~\cite{pdg}.
Agreement is acceptable, but an improvement in the experimental error would allow a much more stringent test of QCD. 

The righthand plot of Figure~\ref{fig:vR} shows results for 
the 4th, 6th, 8th and 10th time-moments of the charmonium vector 
correlator. These moments test the short time behaviour of 
the correlator, where the large time behaviour gives the 
properties of the ground-state charmonium vector meson, 
the $J/\psi$, discussed above. The continuum limit of the 
time-moments can be compared to $q^2$-derivative moments 
of the charm quark vacuum polarisation function derived from 
experimental results for $R_{e^+e^-}$~\cite{kuhn}. Good agreement 
with the experimental results is seen with errors at 1.5\%.

\end{document}